\documentclass[prl,twocolumn,showpacs,preprintnumbers,amsmath,amssymb]
{revtex4}

\usepackage{graphicx}
\usepackage{amsmath}
\usepackage{amsbsy}
\usepackage{dcolumn}
\usepackage{bm}

\begin{document}

\newcommand{\tl}{t_{L}}
\newcommand{\tr}{t_{R}}
\newcommand{\w}{\omega}
\newcommand{\W}{\Omega}
\newcommand{\e}{\varepsilon}
\newcommand{\ed}{\varepsilon_{d}}
\newcommand{\up}{\uparrow}
\newcommand{\down}{\downarrow}
\newcommand{\wph}{\omega_{0}}
\newcommand{\sign}{\text{sign}}
\newcommand{\s}{\sigma}
\newcommand{\RE}{\text{Re}}
\newcommand{\IM}{\text{Im}}
\newcommand{\al}{\alpha}
\newcommand{\la}{\lambda}
\newcommand{\g}{\gamma}
\newcommand{\G}{\Gamma}
\newcommand{\Nf}{N(0)}
\newcommand{\kvec}{{\mathbf k}}
\newcommand{\TK}{T_{K}}
\newcommand{\TKK}{T^{\ast}}
\renewcommand{\vec}[1]{{\mathbf #1}}

\title{Vibrational Sidebands and Kondo-effect in Molecular
  Transistors}

\author{Jens Paaske$^{1,2}$ and Karsten Flensberg$^{2}$}

\affiliation{$^1$\mbox{Institut f\"ur Theorie der Kondensierten
    Materie, Universit\"at Karlsruhe, 76128 Karlsruhe, Germany}\\
  $^2${\O}rsted Laboratory, Niels Bohr Institute fAPG,
  Universitetsparken 5, 2100 Copenhagen, Denmark.}  \date{\today}


\begin{abstract}
  
  Electron transport through molecular quantum dots coupled to a
  single vibrational mode is studied in the Kondo regime. We apply a
  generalized Schrieffer-Wolff transformation to determine the
  effective low-energy spin-spin-vibron-interaction. From this model
  we calculate the nonlinear conductance and find Kondo sidebands
  located at bias-voltages equal to multiples of the vibron frequency.
  Due to selection rules, the side-peaks are found to have strong
  gate-voltage dependences, which can be tested experimentally.  In
  the limit of weak electron-vibron coupling, we employ a perturbative
  renormalization group scheme to calculate analytically the
  nonlinear conductance.

\end{abstract}


\maketitle


In recent years, the study of transport in mesoscopic systems has
branched into investigations of single-electron devices based on
single-molecule transistors~\cite{exp,park00,park02,lian02,yu04}.  Of
particular interest is the possibility of combining electronics with
mechanics, such that the vibrational or configurational modes of the
molecule are coupled to its charge state. A number of interesting
issues have already been addressed in this new field of
nano-electro-mechanics. Firstly, it was shown by Park \textit{et
  al.}~\cite{park00} that quantum mechanical behavior of the center of
mass oscillation of a C$_{60}$ can be excited by the tunneling
electrons and a series of assisted steps were observed in the current.
Similar structures have later been observed in a number of other
experiments.  These steps were associated with simple Franck-Condon
physics, which means that the tunneling rates are modified by the
overlaps of the initial and final states of the oscillator.

It is well established~\cite{glazwing,flen03} that {\it single particle}
resonance tunneling is not destroyed by the electron-vibron coupling,
but instead the resonance breaks up into a number of vibron sidebands.
The question remains, though, whether more intricate {\it many-body}
effects, such as the Kondo resonance, also cooperate with the
electron-vibron coupling to form ``Kondo sidebands''. The usual Kondo
resonance has been observed in several molecular
devices~\cite{park02,lian02} at unusually high temperatures,
and recent experiments~\cite{yu04} on C$_{60}$, and Co based
transistors have revealed marked sideband resonances, which were
suggested to arise from the interplay of a Kondo resonance with a
vibrational mode.

In this letter we demonstrate that, in contrast to sequential
tunneling, which is suppressed by the Franck-Condon overlap
factors, the Kondo resonance remains intact well inside the Coulomb
blockade valley. In fact, the electron-vibron coupling is predicted to
enhance the exchange-coupling and thereby the Kondo-temperature.
Maintaining the quantum coherence of vibrons, we show that the
Kondo-resonance breaks up into a series of vibron sidebands.
Moreover, we demonstrate that parity selection rules prohibit all
sidebands at odd multiples of the oscillator frequency, when tuning
the gate-voltage to the particle-hole symmetric point.


Assuming the energy-level spacing on the molecule to be much
larger than the charging-energy, the system may be described by the
Anderson-Holstein Hamiltonian
\begin{multline}
H=\sum_{\al,\kvec,\s}\xi_{\al\kvec}
  c^{\dagger}_{\al\kvec\s}c_{\al\kvec\s}
 +\ed n_{d}
 + \wph b^{\dagger}b+U n_{d\up}n_{d\down}\\
 +\sum_{\al,\kvec,\s}
   (t_{\al\kvec}d^{\dagger}_{\s}c_{\al\kvec\s}
    + {\rm h.c.})
  +\la\wph(b+b^{\dagger})n_{d}.\label{eq:hamilton}
\end{multline}
where $c^\dagger_{\al\kvec\sigma}$ and $d^\dagger_\sigma$ are creation
operators for electrons in the left and right conduction bands
($\al=L,R$) and on the molecular quantum dot, respectively,
$n_{d\s}=d^{\dagger}_{\s}d_{\s}$, $n_{d}=n_{d\up}+n_{d\down}$ and
$\xi_{\al\kvec}=\e_\kvec-\mu_{\al}$. The vibrational mode of the
molecule is created by the vibron operator $b^{\dagger}$, and $\la$
denotes the dimensionless coupling strength.  Describing the molecule
as a quantum-dot, we have $\ed=(1-2{\cal N})E_{C}$ and $U=2E_{C}$, in
terms of the charging energy $E_{C}$ and the mean occupation number
${\cal N}=C_{g}V_{g}/e$, determined by the gate voltage, $V_{g}$, and
the capacitance to the gate, $C_{g}$.  Unless explicitly stated
otherwise, we shall henceforth work in units where $e=\hbar=k_{B}=1$.

Following Lang and Firsov~\cite{lang63}, the electron-vibron coupling
in the Hamiltonian (\ref{eq:hamilton}) is eliminated by the unitary
transformation $H'=e^{i S_{p}}H e^{-i S_{p}}$, with
$S_{p}=i\la(b-b^{\dagger})n_{d}$:
\begin{multline}\label{eq:hamilton2}
H'=\sum_{\al,\kvec,\s}\xi_{\al\kvec}
  c^{\dagger}_{\al\kvec\s}c_{\al\kvec\s}
 +\ed' n_{d}
 + \wph b^{\dagger}b+U' n_{d\up}n_{d\down}\\
 +\sum_{\al,\kvec,\s}(t_{\al\kvec}e^{\la(b^{\dagger}-b)}
                        d^{\dagger}_{\s}c_{\al\kvec\s}+ {\rm h.c.}),
\end{multline}
where $\ed'=\ed-\la^{2}\wph$ and $U'=U-2\la^{2}\wph$. We now consider
the weak-tunneling limit,
$\G_{\al\kvec}=2\pi\Nf|t_{\al\kvec}|^{2}\ll\min(-\ed',\ed'+U')$, where
$\Nf$ denotes the conduction electron ($ce$) density of states. In
this limit, a generalized Schrieffer-Wolff transformation, devised by
Sch\"{u}ttler and Fedro~\cite{schu88}, may be used to eliminate all
first order terms in $t_{\al}$. To this end, we introduce the
generator
\begin{equation}
S_{v}=i\!\!\!\sum_{\al,\kvec,\s,\eta}(t_{\al\kvec}\zeta_{\al\kvec\s\eta}
n^{\eta}_{d\overline{\s}}d^{\dagger}_{\s}c_{\al\kvec\s}-{\rm h.c.}),
\end{equation}
where $n^{\eta}_{d\overline{\s}}=(1-\eta)/2+\eta n_{d\overline{\s}}$,
with $\eta=\pm 1, \overline{\s}=-\s$, and
$\zeta_{\al\kvec\s\eta}=i\int_{0}^{\infty}\!dt\,
e^{-i(E_{\al\kvec\eta}-i0_{+})t}e^{-A(t)}$, with
$E_{\al\kvec\eta}=\xi_{\al\kvec}-\ed'-(1+\eta)U'/2$ and
$A(t)=\la(e^{-i\wph t}b-e^{i\wph t}b^{\dagger})$.  Applying the
transformation $H''=e^{i S_{v}}H' e^{-i S_{v}}$ and expanding to
second order in $t_{\al}$, one finds that $H''=H_{0}''+H_{\rm
  spin}+H_{\rm dir}+H_{\rm pair}$. We neglect the renormalization of
the kinetic energy term in $H_{0}''$ and, restricting to the regime of
single occupancy, i.e. ${\cal N}\approx 1$ and $\la\wph\ll E_{C}$
($U'>0$), $H_{\rm pair}$ vanishes.  The potential scattering term,
$H_{\rm dir}$, is omitted since it leads to no logarithmic
singularities, and altogether we obtain the effective Hamiltonian
\begin{multline}\label{eq:effhamilton}
H''=\sum_{\al,\kvec,\s}\xi_{\al\kvec}
  c^{\dagger}_{\al\kvec\s}c_{\al\kvec\s}+\wph b^{\dagger}b\\
+\sum_{\al,\kvec,\s;\al',\kvec',\s'}\!\!\!\!\!\!
{\Bbb J}_{\al,\kvec;\al',\kvec'}\,
{\bf S}\cdot c^{\dagger}_{\al'\kvec'\s'}
\frac{{\boldsymbol \tau}_{\s'\s}}{2}c_{\al\kvec\s},
\end{multline}
where ${\bf S}=\frac{1}{2}d^{\dagger}_{\s'}{\boldsymbol
  \tau}_{\s'\s}d_{\s}$ denotes the local spin-$1/2$, and ${\Bbb
  J}_{\al,\kvec;\al',\kvec'}=t_{\al'\kvec'}^{\ast}t_{\al\kvec}
\left[(X_{\al\kvec}^{-}-X_{\al\kvec}^{+})
     +(X_{\al'\kvec'}^{-}-X_{\al'\kvec'}^{+})^{\dagger}\right]$
with
$X_{\al\kvec}^{\eta}=i\!\int_{0}^{\infty}\!\!\!dt\,
e^{-i(E_{\al\kvec\eta}-i0_{+})t+i\eta\la^{2}\sin(\wph t)}
e^{A(0)-A(t)}.$

In this effective Kondo-model, the exchange-coupling ${\Bbb J}$
incorporates the dynamics of the vibron through the displacement
operator $e^{A}$. In the vibron number-state basis it is therefore
convenient to introduce Franck-Condon factors $f_{n'n}=\langle
n'|e^{-A(0)}|n\rangle$~\footnote{$f_{n'n}=\frac{e^{-\la^{2}/2}}
  {\sqrt{n'!\,n!}}  \left[\mathrm{sign}(n-n')\la\,\right]^{|n'-n|}\!
  L^{|n'-n|}_{\min(n',n)}(\la^{2})$} , which allows us to write the
matrix-elements of ${\Bbb J}$ in the more transparent form:
\begin{eqnarray}\label{eq:exch}
\lefteqn{J^{n'n}_{\al',\kvec';\al,\kvec}
\equiv\langle n'|{\Bbb J}_{\al',\kvec';\al,\kvec}|n\rangle=}\\
&&\!\!\!t_{\al'\kvec'}^{\ast}t_{\al\kvec}
\sum_{m=0}^{\infty}
\left\{f_{mn'}f_{mn}\!\left[\,
\frac{1}{\xi_{\al\kvec}-\e_{-}+(m-n')\wph}
\right.\right.\nonumber\\
&&\left.\hspace{34mm}
+\frac{1}{\xi_{\al'\kvec'}-\e_{-}+(m-n)\wph}\right]\nonumber\\
&&\hspace{19mm}
-f_{n'm}f_{nm}\left[\frac{1}{\xi_{\al\kvec}-\e_{+}-(m-n)\wph}\right.
\nonumber\\
&&\left.\left.\hspace{34mm}
+\frac{1}{\xi_{\al'\kvec'}-\e_{+}-(m-n')\wph}\right]\right\},
\nonumber
\end{eqnarray}
valid for
$\xi_{\al\kvec},\xi_{\al'\kvec'},n\wph,n'\wph\ll\min(\e_{+},-\e_{-})$,
where $\e_{-}=\ed'$ and $\e_{+}=\ed'+U'$ are the energies of
intermediate, empty, or doubly occupied states. In this sum, the
energies of intermediate vibron states $|m\rangle$ shift the energy
denominators and the Franck-Condon factors determine the overlap
between initial and final vibron states with intermediate states of
the oscillator shifted by $\sqrt{2}\la\ell_{0}$, where $\ell_{0}$ is
the characteristic oscillator-length.

Since $\sum_{m=0}^{\infty}f_{n'm}f_{nm}=\delta_{n'n}$ and
$f_{n'n}\to\delta_{n'n}$ for $\la\to0$, the usual exchange-coupling,
$J_{\al'\al}=4t_{\al'}^{\ast}t_{\al}/E_{C}$, is recovered in either of
the limits $\wph\to 0$ or $\la\to 0$. More generally, $J^{n'n}$ may be
represented as an asymptotic power series as $E_{C}/\wph\to\infty$,
with leading terms $J^{n'n}_{\al'\al}\propto
J_{\al'\al}(\la\wph/E_{C})^{|n'-n|}$.  In terms of the incomplete
Gamma function, $\gamma(\al,x)$, one has
$J^{00}_{\al'\al}=J_{\al'\al}\,e^{-\la^{2}}(E_{C}/\wph)\!
\sum_{\eta=\pm}\!(-\la^{2})^{\eta \e_{\eta}/\wph}\, \gamma(\eta
\e_{\eta}/\wph,-\la^{2})$, or simply $J^{00}\approx
J_{\al'\al}[1+(\la\wph/E_{C})^{2}]$ for $\la\wph\ll E_{C}$ and ${\cal
  N}=1$, as found earlier in Ref.~\cite{schu88}.  Staying well inside
the Kondo-regime, any finite $\la$ thus leads to a slight {\it
  enhancement} of $J^{00}$, and thereby of the associated Kondo
temperature, $T_{K}\sim D e^{-1/\Nf J^{00}}$ ($2D$ being the $ce$
bandwidth). This is essentially due to the vibron induced reduction of
$|\varepsilon_{\pm}|$. Note also that the (cotunneling) amplitude
$J^{00}$ involves only {\it virtual} shifts of the oscillator and
therefore no Franck-Condon overlap reduces its magnitude. This is in
sharp contrast to the resonant (sequential) tunneling amplitude,
relevant outside the Coulomb-blockade regime, which involves {\it
  real} excitations of the oscillator and is thereby reduced by a
factor of $|f_{00}|^{2}$~\cite{park00,flen03}.

\begin{figure}[t]
\hspace*{2.5mm}
\includegraphics[width=6.0cm]{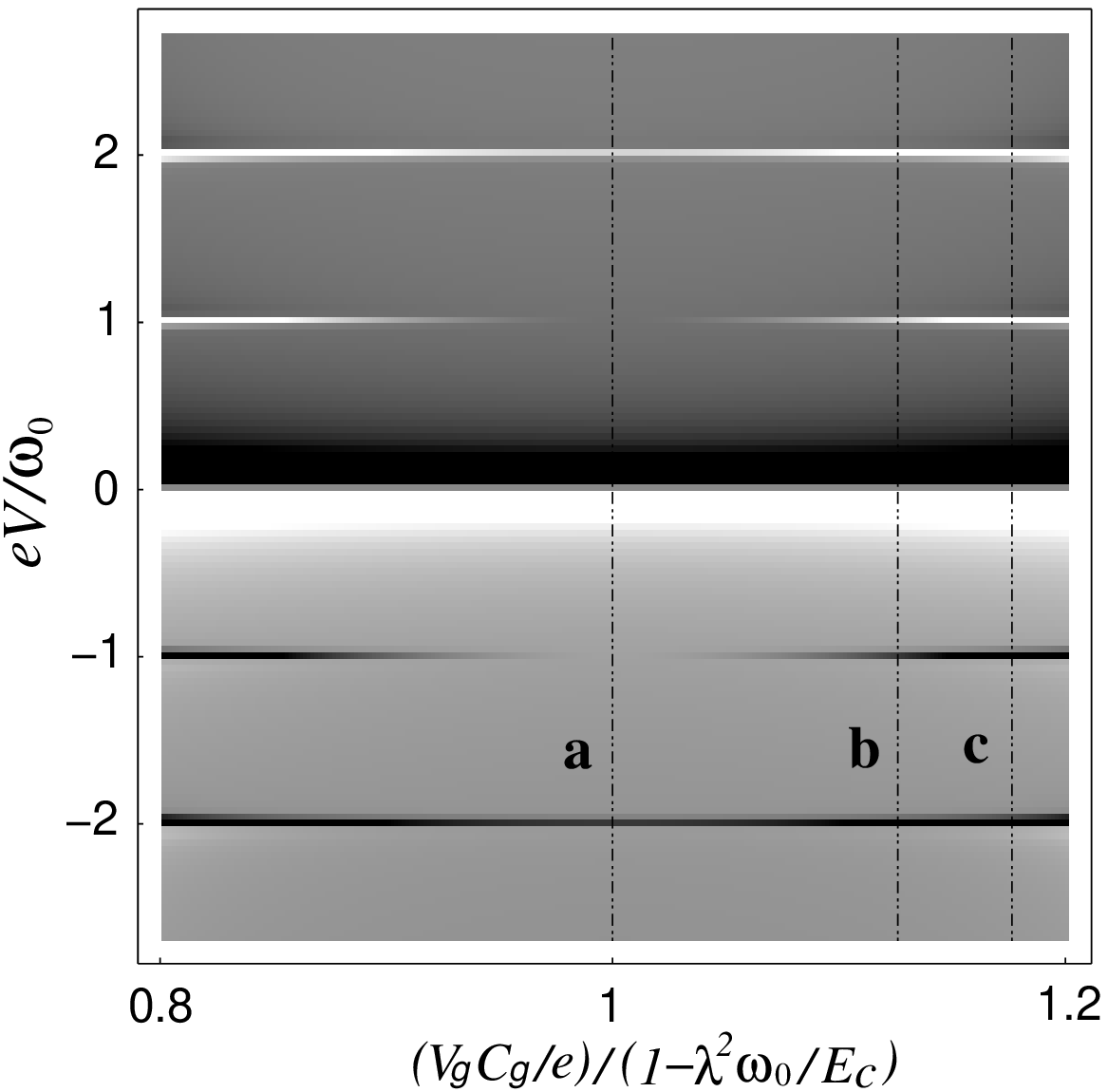}\\\vspace*{2mm}
\includegraphics[width=6.4cm]{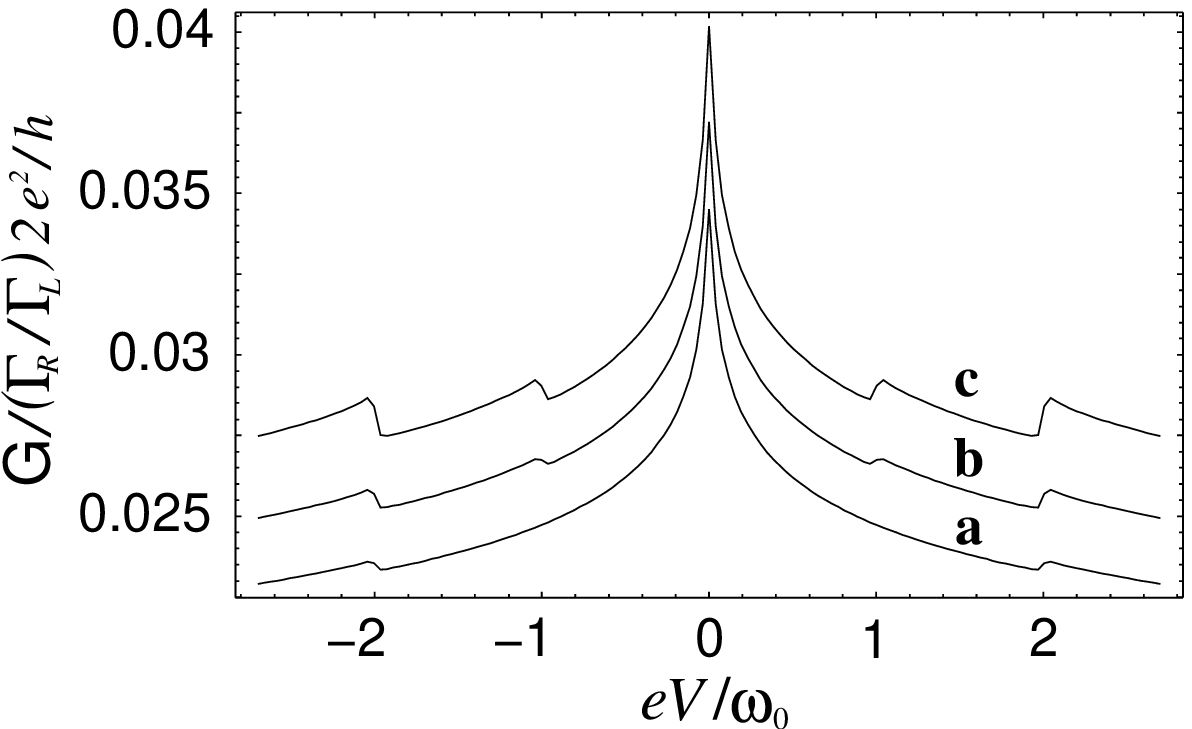}\\\vspace*{2mm}
\caption{\label{fig:cond}
  Upper panel: $\partial^{2}I/\partial V^{2}$ vs. bias and gate
  voltage, for $\la^{2}=3$, $\Nf|t_{L}|^{2}=0.1 \wph$, $D=E_{C}=8
  \wph$, and $T=0.01 \wph$. Black/white indicates large
  negative/positive values. Lower panel: Conductance vs. bias voltage
  for three values of $V_{g}$ corresponding to the vertical black
  lines ({\bf a},{\bf b},{\bf c}) in the upper panel. The lower curve
  (a) corresponds to the ph-symmetric point ${\cal
    N}=1-\la^{2}\wph/E_{C}$.}
\end{figure}

We now consider the case of strongly asymmetric and momentum
independent tunneling amplitudes, $\G_{L}\gg \G_{R}$. The current
traversing the molecule from left to right is then given simply
as~\cite{meir92}
\begin{equation}
I=-\frac{2e}{h}\G_{R}\sum_{\s}
  \int\! d\e\left[f_{L}(\e)-f_{R}(\e)\right]
{\rm Im}{\cal G}^{R}_{\s\s}(\e).\label{eq:current}
\end{equation}
From the equations of motion for the Hamiltonian (\ref{eq:hamilton}),
the local density of states is found to be related to the $ce$
T-matrix as $\IM[{\cal G}_{\s}^{d,R}(\w)]=|t_{\al}|^{-2}\IM[{\rm
  T}_{\al\al}^{\s\s}(\w)]$, and the latter can now be obtained using
the effective Hamiltonian (\ref{eq:effhamilton}). To third order in
$J_{\al,\al'}$, we find that
 \begin{eqnarray}\label{eq:Tmatrix}
 \lefteqn{\hspace{-3mm}{\rm Im}\left[{\rm T}_{\al'\al}^{\s'\s}(\W)\right]
 =-\delta_{\s'\s}\frac{3\pi}{16}\Nf(1-e^{-\wph/T})(1+e^{-\W/T})}
 \nonumber\\
 &&\hspace{-2mm}\times\sum_{\stackrel{\scriptstyle lmn}{\al_{1}\al_{2}}}
 J^{nm}_{\al'\al_{1}}J^{ln}_{\al_{2}\al}e^{-n\wph/T}[1-f(\W+(n-l)\wph)]
 \nonumber\\
 &&\hspace{-2mm}\times\theta(D-|\W+(n-l)\wph|)
 \left\{
\begin{array}{cc} & \\ & \end{array}\hspace{-4.5mm}
\delta_{ml}\delta_{\al_{1}\al_{2}}\!
 +\Nf J^{ml}_{\al_{1}\al_{2}}\right.\nonumber\\
 &&\hspace{-2mm}\left.\times\left[\ln\left|\frac{D}{\W+(n-m)\wph}\right|
 +\ln\left|\frac{D}{\W+(m-l)\wph}\right|\,\right]\right\},
\end{eqnarray}
with the shorthand notation $\ln|D/x|=\ln|D/\sqrt{x^{2}+T^{2}}|$.  In
the asymmetric limit considered here, we take $\mu_{L}=0$ and bias the
right lead to $\mu_{R}=-V$, leaving the position of the molecular
energy levels unaffected. For $T\ll\wph\ll D$, the differential
conductance is then obtained from Eq.~(\ref{eq:current}) as
$G(V)=-\frac{2e^{2}}{\hbar}(\G_{R}/\G_{L})\Nf\sum_{\s}\IM[{\rm
  T}_{LL}^{\s\s}(eV)]$.  From Eq.~(\ref{eq:Tmatrix}), the differential
conductance appears to diverge as $\ln(D/T)$ at voltages corresponding
to multiples of the oscillator frequency, reflecting the onset of a
Kondo-effect assisted by coherent vibron-exchange.  In
Fig.~\ref{fig:cond}, the upper panel shows a gray-scale plot of
$\partial^{2}I/\partial V^{2}$ as a function of bias-voltage $V$ and
mean occupation number (gate-voltage) ${\cal N}=C_{g}V_{g}/e$. The
lower panel shows three cuts revealing the side-band resonances on the
flanks of the central zero-bias resonance.

By tuning the gate-voltage to ${\cal N}=1-\la^{2}\wph/E_{C}$ one
reaches the particle-hole (ph) symmetric point where $\e_{+}=-\e_{-}$,
and using the general symmetry $f_{n'n}=(-1)^{|n'-n|}f_{nn'}$ one
finds from Eq.~(\ref{eq:exch}) that $J^{n'n}\propto[1+(-1)^{|n'-n|}]$,
implying that all spin-exchange processes involving the emission or
absorption of an {\it odd} number of vibrons are prohibited at this
particular gate-voltage. This parity selection rule, reflecting the
inversion symmetry of the Kondo-Hamiltonian (\ref{eq:effhamilton}) at
low energies, has important experimental bearings, since it predicts
that all side-band resonances in the differential conductance located
at voltages equal to {\it odd} multiples of $\wph$, must vanish when
tuning the gate-voltage to the symmetric point, corresponding to the
midpoint of the Coulomb-blockade valley. This is apparent in
Fig.~\ref{fig:cond}, where the conductance peak at $V=\wph$ disappears
as ${\cal N}$ approaches the symmetric point (cf. curve {\bf a}).
Note, however, that any appreciable vibron modulation of the tunneling
amplitudes will break the inversion symmetry\cite{corn04} and thereby
destroy this selection rule.


The logarithmic divergences appearing in third order perturbation
theory call for a resummation of leading logarithmic contributions to
all orders. This is done using the perturbative renormalization group
(RG) method for frequency dependent couplings developed in
Ref.~\cite{rosc03}. Parametrizing the dimensionless couplings,
$g_{n'n}=\Nf J^{n'n}_{LL}$, by the total energy of the ingoing
conduction electron and vibron-state, we arrive at the infinite
hierarchy of coupled (1-loop) RG-equations:
\begin{eqnarray}
\lefteqn{\hspace{-7.2mm}\frac{\partial g_{n'n}(\w)}{\partial\ln D}
=-\frac{1}{2}\!\!\sum_{m=0}^{\infty}\!
\left[
g_{n'm}(0)g_{mn}((m-n)\wph)\Theta_{\w+(n-m)\wph}\right.}\nonumber\\
&&\hspace{5mm}\left.+g_{n'm}((n'-m)\wph)g_{mn}(0)\Theta_{\w+(m-n')\wph}
\right],\label{eq:RGeq}
\end{eqnarray}
with $\Theta_{\w}=\Theta(D-|\w|)$. We shall restrict our attention to
the ph-symmetric point and assuming that $\la\wph\ll E_{C}$ we may
truncate this hierarchy and consider merely the lowest four coupled
equations involving $g_{00}$, $g_{02}=g_{20}$ and $g_{22}$. The
solution to this reduced set of equations is characterized by the
parameters
\begin{equation}
\delta=
\frac{g_{22}-g_{00}}{g_{00}g_{22}-g_{20}^{2}},\,\,\,\,\,\,
\al=\frac{2 g_{20}}{g_{00}g_{22}-g_{20}^{2}}
\end{equation}
and $\TK=D
e^{-2/(g_{00}+g_{22}+\sqrt{(g_{22}-g_{00})^{2}+4g_{20}^{2}})}$, where
$g_{n'n}\equiv g_{n'n}(D;2n\wph)$. All three parameters are invariant
under the perturbative RG-flow from the initial cut-off $D_{0}$ down
to $D=2\wph$.  At scale $D_{0}$, we have $\al,\delta\sim
(\la\wph/E_{C})^{2}/g_{00}$, and therefore our truncation of
Eq.~(\ref{eq:RGeq}) remains valid throughout the RG-flow roughly when
$\max(\al,\delta)/\ln(T/\TKK)\ll 1$ (see below). Staying within the
perturbative regime, we assume that $\wph\gg T\gg\TK$.
 
We first solve the RG-eqs. for the constant coefficients $g_{n'n}$ and
the frequency dependent renormalized couplings are then obtained
simply by integrating Eqs.~(\ref{eq:RGeq}):
\begin{eqnarray}
g_{00}(\w)&=&\frac{1}{\ln(|\w|/\TKK)}
+\frac{\al^{2}}{8\ln^{2}(2\wph/\TKK)}\label{eq:g00}\\
&&\hspace*{-12mm}\times\!\sum_{\nu=0,1}\!\!\Theta_{\nu}(\w)\!
\left(\frac{1}{\ln(|\w-2\nu\wph|/\TKK)}
-\frac{1}{\ln(2\wph/\TKK)}\right)\nonumber\\
g_{20}(\w)&=&\!\frac{\al}{4}\!\sum_{\nu=0,1}
\left[\frac{1}{\ln^{2}(\max(2\wph,|\w-2\nu\wph|)/\TKK)}\right.
\label{eq:g20}\\
&&\left.\hspace*{-15mm}+\frac{2\Theta_{\nu}(\w)}{\ln(2\wph/\TKK)}
\left(\!\frac{1}{\ln(|\w-2\nu\wph|/\TKK)}
-\frac{1}{\ln(2\wph/\TKK)}\right)\right],\nonumber
\end{eqnarray}
with $\TKK=\TK(\TK/\wph)^{(\sqrt{\al^{2}+\delta^{2}}+\delta)/
  (2\ln(2\wph/\TK)+\sqrt{\al^{2}+\delta^{2}}-\delta)}$ and
$\Theta_{\nu}(\w)=\Theta(2\wph-|\w-2\nu\wph|)$. Note that in Eqs.
(\ref{eq:g00},\ref{eq:g20}) we have retained only terms which will
contribute to order $\max(\al,\delta)^{2}/\ln^{2}(T/\TKK)$ in the
conductance.

While the logarithmic singularities at $\w=0$ are cut off by
temperature, those at $\w=2\wph$ will instead be contained by
$\sqrt{T^{2}+\g^{2}}$, with $\g$ given by the transition rate from
vibron-state $|2\rangle$ to $|0\rangle$.  Using the golden rule with
the renormalized coupling $g_{20}(\w)$, we find $\g\approx
\pi\wph\al^{2}/(4\ln^{2}(2\wph/\TK))$. Similarly, the broadening of
the vibron states, induces a broadening of the step-functions in
Eqs.~(\ref{eq:g00},\ref{eq:g20}) by $T$ or $\sqrt{T^{2}+\g^{2}}$ for
steps near $\w=0$ or $\w=2\wph$, respectively.

\begin{figure}[t]
\hspace*{-4mm}
\includegraphics[width=7.cm, height=5cm]{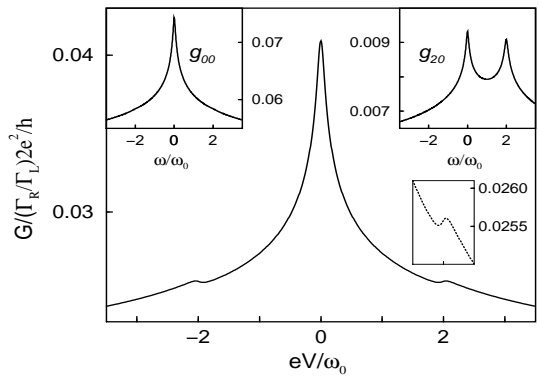}
\vspace*{-3mm}
\caption{\label{fig:rgcombi}
  Conductance versus bias voltage using {\it perturbatively}
  renormalized couplings. Gate voltage is tuned to the $ph$-symmetric
  point and $\al=4.30$, $\delta=7.54$, $\TK=1.51\times 10^{-4}\wph$,
  $\TKK=7.18\times 10^{-8}\wph$, and $T=0.05\wph$, i.e.
  $\al/\ln(T/\TKK)=0.32$ and $\delta/\ln(T/\TKK)=0.56$, corresponding
  to bare parameters: $\la^{2}=4.5$, $\Nf|t_{L}|^{2}=0.1 \wph$ and
  $D=E_{C}=8\wph$. Insets show the renormalized couplings $g_{00}(\w)$
  and $g_{20}(\w)$, as well as a zoom in the conductance curve showing
  the satellite peak on a separate conductance scale but on same voltage
  scale.}
\end{figure}
The renormalized conductance is now obtained by inserting in the
formula
$G(V)=(2e^{2}/h)(\G_{R}/\G_{L})(3\pi^{2}/4)$\\$[g_{00}(eV)^{2}+
\sum_{\nu=\pm}\Theta(\nu eV-2\wph)g_{20}(\nu eV)^{2}]$, and indeed
when expanding this result in bare couplings, we recover the result
obtained by expanding Eq.~(\ref{eq:Tmatrix}) to order
$(\la\wph/E_{C})^{4}$.  Including the broadening in both couplings and
$\Theta$, the result is plotted in Fig.~\ref{fig:rgcombi}. This figure
bears a certain resemblance to the curve obtained in
Ref.~\cite{koni96} at $T\ll\TK$, using an entirely different method.
  
The possible decoherence effects which will arise when coupling the
oscillator to phonons within the leads remain an open question.  It is
straightforward to generalize the Schrieffer-Wolff transformation
applied here to a system where the molecule-oscillator is coupled to a
separate bath of oscillators. However, even determining the effects on
the leading logarithms for a given Q-factor involves a rather involved
cumulant expansion. We expect the Kondo-effect to be more pronounced
when dealing solely with {\it intra}-molecular vibrations, since these
have been demonstrated to have particularly large
Q-factors~\cite{gure98}.

In the case of nearly symmetric couplings, we can no longer assume the
oscillator to be in equilibrium with the conduction electrons of one
specific side of the junction\cite{nonequil}. In line with the
findings of Ref.~\cite{koni96,aji03}, we expect that nonequlibrium
effects may in fact serve to enhance the Kondo side-peaks.

In conclusion, we have demonstrated the viability of an inelastic
Kondo-effect carried by coherent vibron-assisted exchange tunneling,
which can be observed as Kondo-sidebands in the nonlinear conductance.
In contrast to the case of an applied microwave field~\cite{koga04},
the zero-bias resonance is {\it not} suppressed by the vibronic
coupling, and it may therefore be difficult to discern the satellites
from the background conductance.  Nevertheless, even with very weak
satellites, it should be possible to track their dependence on $V_{g}$
(possibly in a plot of $\partial^{2}I/\partial V^{2}$) and thereby
test our prediction that satellites at odd multiples of $\wph$ are
strongly reduced at the ph-symmetric point. Faint sidebands to a
zero-bias Kondo peak have indeed been observed in the recent
experiment by Yu {\it et al.}\cite{yu04}. In contrast to the findings
reported here, however, both satellite {\it peaks} and {\it dips} were
observed in the nonlinear conductance.


We thank P. Brouwer, A. Rosch, P. Sharma and P. W\"{o}lfle for useful
discussions. This research was supported by the Center for Functional
Nanostructures (J.~P.) and by the European Commission through project
FP6-003673 CANEL of the IST Priority\footnote{The views expressed in
  this publication are those of the authors and do not necessarily
  reflect the official European Commission's view on the subject.}.


\end{document}